\documentclass[a4paper,11pt]{article}

\usepackage{pos}
\usepackage{caption}
\usepackage{subcaption}
\usepackage{lipsum} 

\title{Cross Correlation of IceCube Neutrinos with Tracers of Large Scale Structure
}

\ShortTitle{Cross Correlation of IceCube Neutrinos with Tracers of Large Scale Structure
}

\author{The IceCube Collaboration \\{\normalsize \normalfont(a complete list of authors can be found at the end of the proceedings)}\\}

\emailAdd{david.guevel@icecube.wisc.edu}
\emailAdd{kefang@physics.wisc.edu}

\abstract{

 The origin of most astrophysical neutrinos is unknown, but extragalactic neutrino sources may follow the spatial distribution of the large-scale structure of the universe.
 Galaxies also follow the same large scale distribution, so establishing a correlation between galaxies and IceCube neutrinos could help identify the origins of the diffuse neutrinos observed by IceCube.
 Following a preliminary study based on the WISE and 2MASS catalogs \cite{fangCrosscorrelationStudyHighenergy2020}, we will investigate an updated galaxy catalog with improved redshift measurements and reduced stellar contamination.
 Our IceCube data sample consists of track-like muon neutrinos selected from the Northern sky.
 The excellent angular resolution of track-like events and low contamination with atmospheric muons is necessary for the sensitivity of the analysis.
 Unlike a point source stacking analysis, the calculation of the cross correlation does not scale with the number of entries in the catalog, making the work tractable for catalogs with millions of objects.
 We present the development and performance of a two-point cross correlation of IceCube neutrinos with a tracer of the large scale structure.

\vspace{4mm}
{\bfseries Corresponding authors:}
David Guevel$^{1*}$, Ke Fang$^{1}$\\
{$^{1}$ \itshape  Dept. of Physics and Wisconsin IceCube Particle Astrophysics Center}\\[4mm]
$^*$ Presenter

\ConferenceLogo{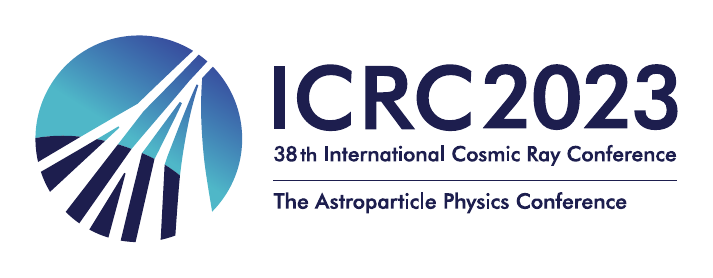}

\FullConference{The 38th International Cosmic Ray Conference (ICRC2023)\\ 26 July -- 3 August, 2023\\ Nagoya, Japan}
}

\begin{document}

\maketitle

\section{Introduction}\label{intro}
IceCube has observed high energy neutrinos with an astrophysical origin \cite{collaboration*EvidenceHighEnergyExtraterrestrial2013}; however, the astrophysical sites where the majority of them are produced are not known.
NCG 1068 and TXS 0506+056 have been identified as neutrino source candidates \cite{icecubecollaborationNeutrinoEmissionDirection2018a, icecubecollaborationEvidenceNeutrinoEmission2022}, but they cannot account for more than a few percent of the observed diffuse neutrino flux.
NGC 1068 is a Seyfert galaxy with a high star formation rate whose neutrino emission is 1-2 orders of magnitude brighter than what would be predicted from its $\gamma$-ray emission\cite{icecubecollaborationEvidenceNeutrinoEmission2022}.
Previous searches have focused on the connection between $\gamma$-ray and neutrino production \cite{abbasiSearchCorrelationsHighenergy2023, abbasiSearchMultiflareNeutrino2021, aartsenIceCubeSearchHighenergy2020}, but the excess neutrino emission from NGC 1068 suggests that the $\gamma$-ray emission may not be the most effective tracer of neutrino production.
If the majority of neutrino sources resemble star forming Seyfert galaxies like NGC 1068, whose bolometric luminosity peaks in the infrared, then a catalog of galaxies found by an infrared survey can be a powerful tracer of neutrino production.
Moreover, whether the sites of neutrino production are infrared galaxies or another astrophysical source, both neutrinos and galaxies will likely trace the same underlying large-scale mass distribution.
Infrared galaxies selected from the 2MASS Redshift Survey \cite{huchra2MASSREDSHIFTSURVEY2012} have been used to constrain IceCube neutrino sources \cite{sclafaniCorrelationIceCubeNeutrinos2019, aartsenConstraintsNeutrinoEmission2020}.
The 2MASS Redshift Survey obtained spectroscopic redshift for a relatively small ($\sim ~45000$) sample of bright 2MASS galaxies.
We take the complementary approach by using a large sample of galaxies with poorly known redshift.

The two-point cross correlation is a commonly used tool in cosmology for detecting anisotropy.
The cross correlation is effective for large catalogs because the computation does not scale with the number of catalog members, unlike likelihood-based point source searches.
We are developing a search for neutrino sources using a catalog of infrared  galaxies observed by WISE and 2MASS \cite{wrightWIDEFIELDINFRAREDSURVEY2010a, skrutskieTwoMicronAll2006}.
We will cross correlate this galaxy catalog with a selection of muon neutrinos from the northern sky ($\delta > -10^\circ$), where the atmospheric muon background is reduced.
There are two dominant sources of background events in track-like muon neutrino samples.
The first are tracks created by muons created in the atmosphere.
This background is effectively eliminated by selecting only events which travel through the Earth before reaching IceCube.
The second background is caused by muon neutrinos created in the atmosphere.
These events cannot be distinguished from neutrinos with an astrophysical origin on an event-by-event basis.
The atmospheric neutrinos follow will follow an approximately isotropic distribution, which will be distinguishable in the cross correlation from the anisotropic distribution of the galaxies in our catalog.
We have developed a Python module called \texttt{nuXgal} to perform this analysis, which is available on GitHub \footnote{https://github.com/dguevel/nuXgal}.
Analyses based on auto-correlation and cross-correlation have been performed before by the IceCube Collaboration.
An auto-correlation analysis found no statistically significant evidence for anisotropy in the diffuse neutrino emission \cite{aartsenSearchesSmallscaleAnisotropies2015, Glauch:2017kQ}.
A cross correlation analysis of IceCube neutrinos and the unresolved {\it Fermi}-LAT background found no significant correlation and constrained the neutrino emission from unresolved blazars to less than 1\% of the observed neutrino flux \cite{negroCrosscorrelationStudyIceCube2023}.

\section{Galaxy Catalog}\label{catalog}
Star forming galaxies are bright in infrared emission originating from thermal emission of interstellar dust.
We constructed a galaxy sample using Wide Field Infrared Explorer (WISE) and Two Micron All Sky Survey (2MASS) observations.
As part of its extended mission, WISE has observed the full sky in the mid-infrared at 3.4$\mu m$ (W1) and 4.6$\mu m$ (W2).
WISE observations have been reprocessed with improved photometry and astrometry to produce a catalog called unWISE \citep{schlaflyUnWISECatalogTwo2019}.
The unWISE catalog improves the depth and completeness of previous WISE catalogs and contains over two billion sources.
2MASS observed the full sky in the near-infrared at 1.25$\mu m$ (J), 1.65$\mu m$ (H), and 2.16$\mu m$ (K$_s$).
The 2MASS Point Source Catalog contains 471 million sources.

\begin{figure}
\centering
\begin{subfigure}[c]{0.5\textwidth}
  \centering
  \includegraphics[width=0.9\textwidth]{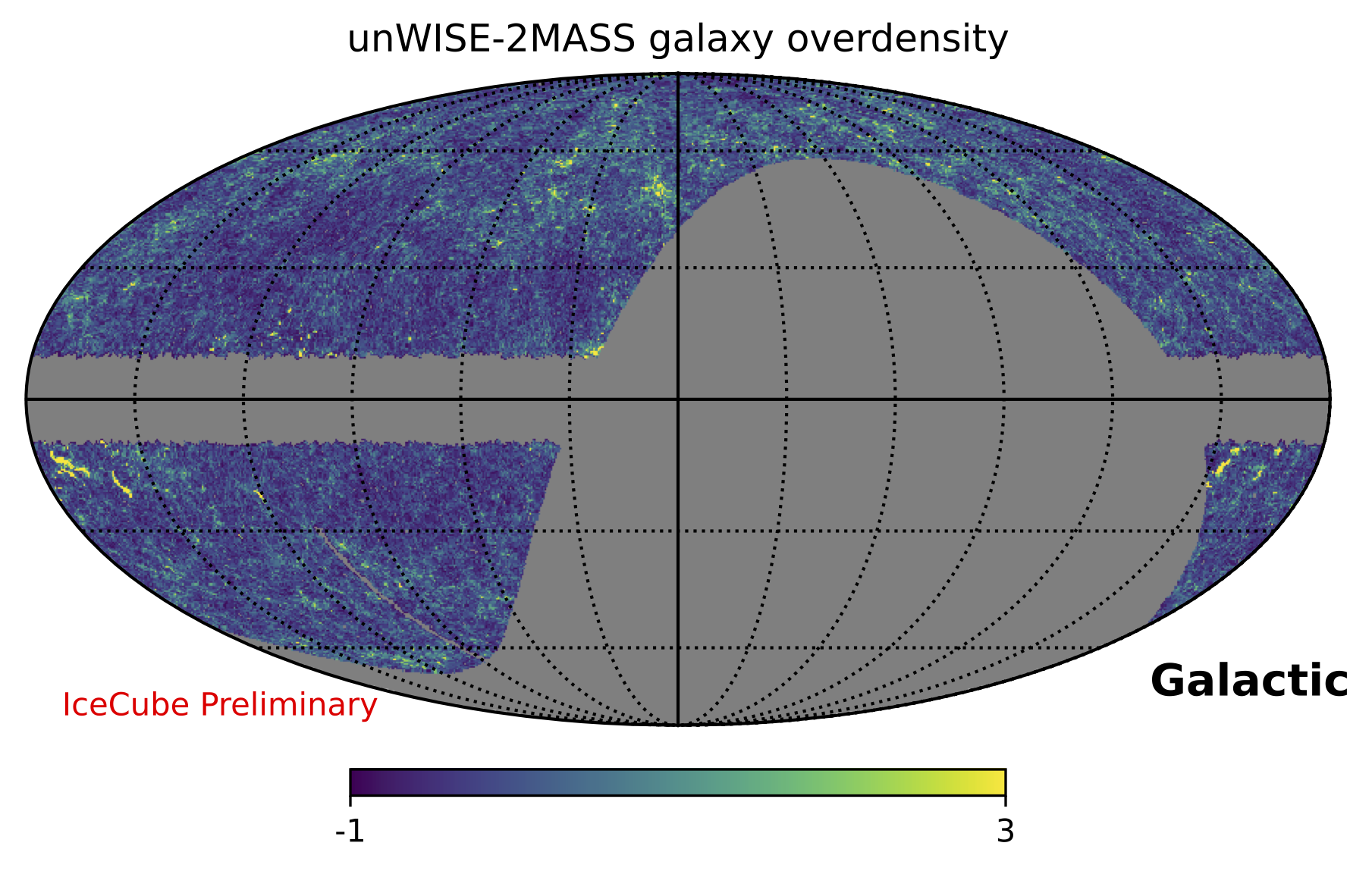}
\end{subfigure}%
\begin{subfigure}[c]{0.5\textwidth}
  \centering
  \includegraphics[width=1.0\linewidth ]{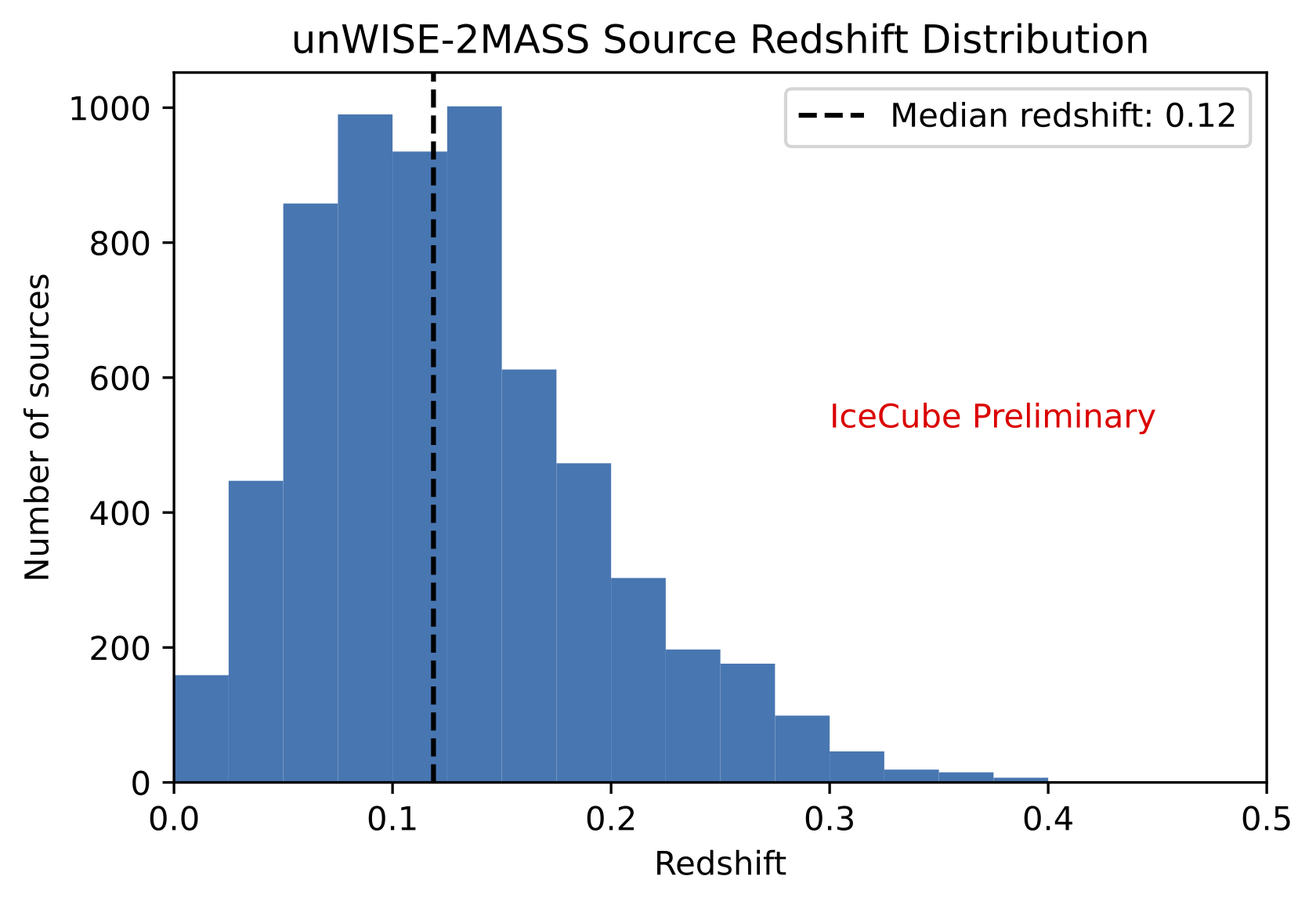}
\end{subfigure}
\caption{Left: Overdensity of unWISE-2MASS sources in the northern sky.
The full sky selection contains approximately 8 million sources after all filtering is applied.
Right: Redshift distribution of unWISE-2MASS sources which have counterparts in the GAMA catalog.
The GAMA fields cover approximately 250 square degrees or less than 1\% of the sky.
In the GAMA fields, 65\% of unWISE-2MASS sources have a GAMA counterpart within 1 arcsecond.
The GAMA cross match sample contains 6349 sources.}
\label{fig:unwise}
\end{figure}

The unWISE catalog contains photometry for only the W1 and W2, the two bands that do not need the coolant that was expended during the initial mission.
The lack of observations at longer wavelength make it challenging to distinguish between stars and galaxies; however, a selection based on WISE and 2MASS color is an effective classifier \cite{kovacsStarGalaxySeparation2015}.
For each source in the unWISE catalog, we identify the nearest source in the 2MASS Point Source Catalog.
The unWISE astrometry has an angular uncertainty of a few arcseconds, so a 2MASS source that is nearer than a few arcseconds from the unWISE source is likely to be the same source.
We adopt a threshold distance of 3 arcseconds.
If the separation is more than this threshold, we exclude the source from our galaxy catalog.
Although it is possible for a 2MASS source to be associated with two unWISE sources under this scheme, the angular density of the unWISE sources make this an negligible contribution to the overall sample.
For a representative high galactic latitude field ($l=90^\circ$, $b=50^\circ$), the angular separation between unwise sources is $27\pm12$ arcminutes.
We exclude sources with $W1 - J > -1.7$ and $J < 16.5$ as suggested by \cite{kovacsStarGalaxySeparation2015} to distinguish between stars and galaxies.
We further exclude sources with $W2 > 15.5$ to preferentially select low redshift sources \cite{schlaflyUnWISECatalogTwo2019}.
The galaxies from the northern sky are shown in Figure \ref{fig:unwise} and the resulting number of sources are summarized in Table \ref{tab:filtertable}.
The unWISE catalog is $\sim 99\%$ complete for $W2 < 15.5$.
The completeness of the same cuts using a shallower WISE catalog is 70.1\%.
The unWISE catalog is deeper and more complete, so the completeness will be at least that large.
Further study of the unWISE-2MASS galaxy catalog will be needed before interpreting our results.

\begin{table}
\begin{center}
\begin{tabular}{|c|c|c|}
\hline
  & Percent of Original Sample & Number of Sources \\  \hline
 Original Sample & 100\% & $\sim 2.0\times 10^9$ \\  
 2MASS Cross Match & 16\% & $\sim 3.2\times 10^8$ \\
 $W2 < 15.5$ & 12\% & $\sim 2.4 \times 10^8$ \\
 $J < 16.5$ & 11\% & $\sim 2.2 \times 10^8$ \\
 $W1 - J < -1.7$ & 0.4\% & $\sim 8.0 \times 10^6$ \\
\hline
\end{tabular}
\caption{Fraction and number of sources remaining after each stage in catalog filtering. The percentages are estimated from a subset of the sky containing $2.5 \times 10^9$ sources.}
\label{tab:filtertable}
\end{center}
\end{table}

We measured the redshift distribution using the Galaxy and Mass Assembly Survey (GAMA) \cite{driverGalaxyMassAssembly2022}.
GAMA is a redshift survey that provides redshifts for more than 300,000 objects across more than 250 square degrees.
We applied our unWISE-2MASS selection to galaxies in the GAMA fields which were located within 1 arcsecond of the unWISE coordinates.
The redshift distribution of the matched galaxies is shown in Figure \ref{fig:unwise}.
The median redshift was 0.12 and the maximum redshift was less than 0.4.

\section{Analysis Method}\label{method}
The cross-correlation is widely used in cosmology to study the large scale structure of the universe.
We perform the cross-correlation between the unWISE-2MASS galaxy catalog and IceCube track-like muon neutrinos with declination greater than -10 degrees in three logarithmically spaced bins between 1 TeV and 1 PeV \cite{icecubecollaborationIceCubeDataNeutrino2021a}.
The two-point cross correlation power spectrum of galaxies ($g$) and neutrinos ($\nu$) in energy bin $i$ can be most easily calculated in the spherical harmonic representation of the two sky maps.
\begin{equation}
    C_{l,i}^{g\nu} = \frac{1}{f_{\rm sky}(2l+1)}\sum_l a^{g*}_{lm} a^\nu_{lm,i}
\end{equation}
where spherical harmonic coefficients of source population are $a_{lm} = \sum_i \delta({\bf x})\ Y^*_{lm}({\bf x})$
and $\delta({\bf x})$ is the overdensity of either neutrinos in each energy bin or galaxies defined as $\delta({\bf x}) = \frac{n({\bf x}) - \bar{n}}{\bar{n}}$ and $\bar{n}$ is the sky-averaged counts.

For a two component neutrino sample, the cross power spectrum can be decomposed into a sum of two components: an astrophysical component originating from a source population that has the same underlying distribution as the galaxy sample, and a background component \citep{fangCrosscorrelationStudyHighenergy2020}
\begin{equation}
    C^{g\nu}_{l,i}= f\ C_{l,i}^{\nu, {\rm astro}} + (1-f)\ C_{l,i}^{\nu, {\rm background}}.
\end{equation}

If neutrinos trace the galaxy catalog, then the parameter $f_i$ represents the fraction of neutrinos that originate from the galaxies, ie., $f_i=\frac{n_{\rm astro}}{n_{\rm total}}$.
The diffuse muon neutrino flux requires that the astrophysical fraction for the three energy bins is at most $\sim 1\%$, $\sim 10\%$, and $\sim 50\%$ in order of increasing energy \cite{abbasiImprovedCharacterizationAstrophysical2022}.

Each $C^{g\nu}_{l,i}$ is normally distributed around the true value,
\begin{equation}
    \log{L(C^{g\nu}_{l,i}|f_i)} = \sum_l \frac{\left(C^{g\nu}_{l,i} - \left( f \langle C_{l,i}^{\nu, {\rm astro}} \rangle + (1-f) \langle C_{l,i}^{\nu, {\rm background}} \rangle \right) \right)^2}{2 \sigma_{l,i}^2}
\end{equation}
where angular brackets indicate the expected value. 
The significance of the cross correlation can be evaluated using the log-likelihood ratio as a test statistic, 
\begin{equation}
    {\rm TS} = 2\left( \log L(\hat{f_i}) - \log L(0)\right)
\end{equation}
where $\hat{f_i}$ is the maximum likelihood estimate constrained such that $\hat{f_i}\ge0$.

The uncertainty $\sigma_{l,i}$ in the cross spectrum depends on the signal purity $f_i$ in a non-trivial way.
We performed MC simulations with varying signal purity to parameterize the uncertainty as a function of signal purity.
The lowest energy bin (1 - 10 TeV) is completely dominated by background events even for signal injection several times greater than the observed astrophysical diffuse muon neutrino flux \citep{stettnerMeasurementDiffuseAstrophysical2019, abbasiImprovedCharacterizationAstrophysical2022}, so the uncertainty is effectively constant up to the maximum $f_i$ allowed by the diffuse muon neutrino flux.
The higher energy bins have higher signal purity, so $\sigma_{l,i}$ does change for the 10 - 100 TeV and 100 - 1000 TeV energy bins.
For these energy bins, we use bootstrap resampling to estimate $\sigma_{l,i}$.
For a set of neutrinos in an energy bin $\{\nu_j: j=0, ..., N\}$, we create 100 new sets of neutrinos by drawing $N$ events from the original set with replacement.
For each resampled neutrino set, we calculate $C^{g\nu}_{l,i}$.
Finally, we use the new set of $C^{g\nu}_{l,i}$ to estimate the standard deviation $\sigma_{l,i}$.
The bootstrapped $\sigma_{l,i}$ agree with Monte Carlo calculations of $\sigma_{l,i}(f_i)$ across the range of $f_i$ allowed by the diffuse muon neutrino fit.
The bootstrap method can be estimated purely from data without knowing the true signal purity.

\begin{figure}
\centering
\begin{subfigure}[c]{0.33\textwidth}
  \centering
  \includegraphics[width=\textwidth]{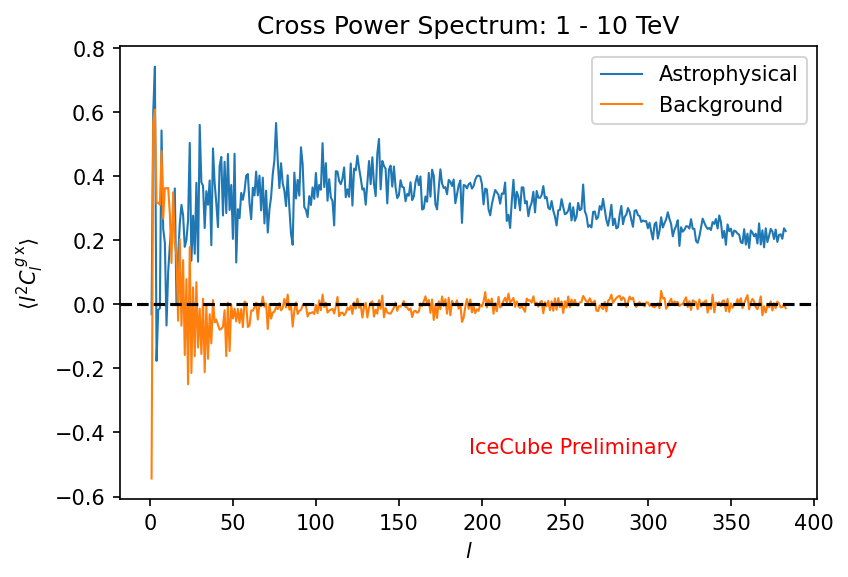}
\end{subfigure}%
\begin{subfigure}[c]{0.33\textwidth}
  \centering
  \includegraphics[width=\linewidth ]{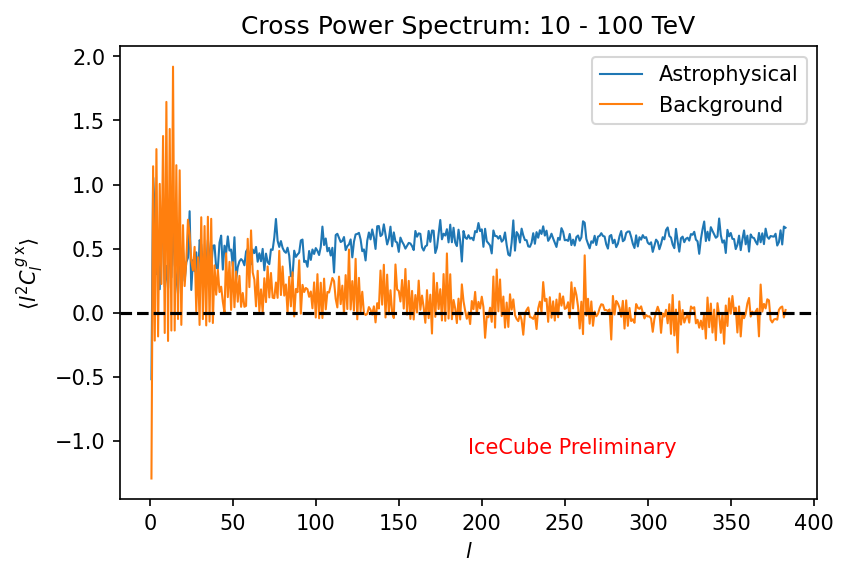}
\end{subfigure}
\begin{subfigure}[c]{0.33\textwidth}
  \centering
  \includegraphics[width=\linewidth ]{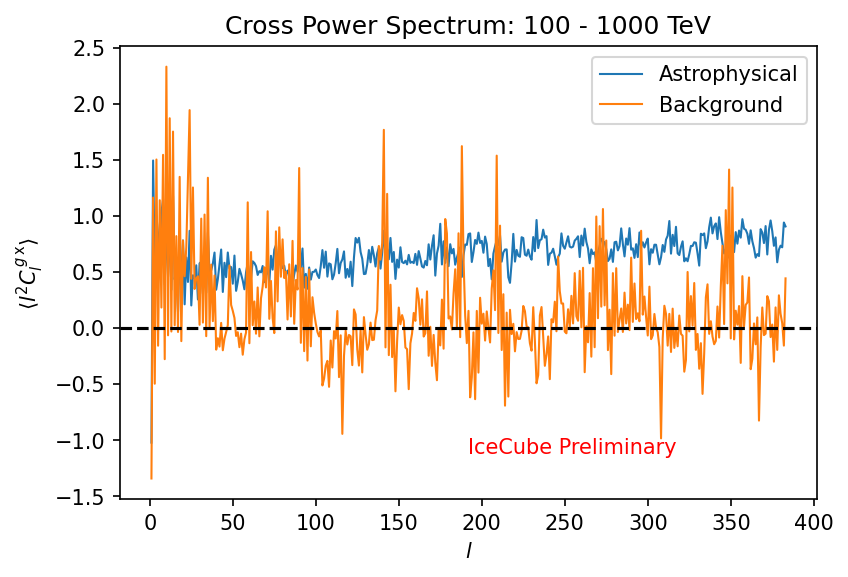}
\end{subfigure}
\caption{Cross power spectra for each energy bin. The blue line is the expected cross power spectrum for a data sample that originates purely from the unWISE-2MASS galaxy sample. The orange line is the cross power spectrum for background only. Both are averaged over 500 realizations.}
\label{fig:cross_power}
\end{figure}

The IceCube point spread function (PSF) and effective area depend on event declination and reconstructed energy.
In principle, a beam function which depends on the declination and is averaged over the energy bin can be used; however, this requires the multiplication of large matrices and is not available in the \texttt{healpy} library.
Given the mature tools that the IceCube collaboration has developed for simulated realistic data sets, we directly compute the expected cross correlation by generating pseudo-trials and taking the mean.
These features are not available in \texttt{healpy} and are computationally challenging.
We simulate 500 purely astrophysical and purely background pseudo-experiments and use these to directly calculate $\langle C_{l,i}^{\nu, {\rm astro}} \rangle$ and $\langle C_{l,i}^{\nu, {\rm background}} \rangle$.
The simulated $\langle C_{l,i}^{\nu, {\rm astro}} \rangle$ and $\langle C_{l,i}^{\nu, {\rm background}} \rangle$ are shown in Figure \ref{fig:cross_power}.

For the purpose of hypothesis testing, we calculate the test statistic distribution in the case of purely background data.
We use internal IceCube tools to generate pseudo-trials and use \texttt{nuXgal} to evaluate the test statistic.
If we do not constrain $f_i>=0$, the test statistic follows a $\chi^2$ distribution with one degree of freedom, as is expected from Wilks' theorem \citep{wilksLargeSampleDistributionLikelihood1938}.
We have verified this with background-only pseudo-experiments.
In the constrained case ($f_i \ge 0$), the test statistic is zero for half the pseudo-experiments and the other half follow a $\chi^2$ distribution with one degree of freedom.
In the unconstrained case, $f_i$ would be normally distributed around zero.
In the constrained case, every pseudo-trial where the maximum likelihood $f_i$ would be less than zero is now equal to zero.
These trials then have ${\rm TS}=0$.
With the background test statistic distribution, we can evaluate the statistical significance of a given pseudo-trial and in the future the significance of unblinded data.

Preliminary work suggests that the cross correlation will be able to constrain the fraction of neutrinos which come from sources that are correlated with the unWISE-2MASS galaxy catalog.
We created sets of pseudo-trials with signal injected up to the level of the observed diffuse muon neutrino flux.
For each set of pseudo-trials, we calculated the fraction of trials with test statistic that exceed the median of the background test statistic distribution 90\% of the time.
We define the sensitivity as the level of signal injection where this threshold is crossed.
The low and intermediate energy bins (1--10 TeV and 10--100 TeV) have sensitivity below the observed diffuse muon neutrino flux which suggests that the cross correlation will be able to constrain the fraction of neutrinos which come from our galaxy catalog.

\section{Conclusion}\label{conclusion}

The two-point cross correlation is a promising technique for the discovery of anisotropy in IceCube neutrinos.
We have developed a galaxy catalog based on WISE and 2MASS observations that traces the large scale mass distribution of the universe.
This galaxy catalog has a median redshift of 0.12 and a maximum redshift of 0.4.
Our implementation of the cross correlation, which is based on a straightforward Gaussian likelihood, is statistically well behaved.
Preliminary work shows that the cross correlation has potential to constrain the sources of the diffuse muon neutrino flux observed by IceCube.

\bibliographystyle{ICRC}
\bibliography{references}

%

\clearpage

\section*{Full Author List: IceCube Collaboration}

\scriptsize
\noindent
R. Abbasi$^{17}$,
M. Ackermann$^{63}$,
J. Adams$^{18}$,
S. K. Agarwalla$^{40,\: 64}$,
J. A. Aguilar$^{12}$,
M. Ahlers$^{22}$,
J.M. Alameddine$^{23}$,
N. M. Amin$^{44}$,
K. Andeen$^{42}$,
G. Anton$^{26}$
% ...
\\
\\
% ...
$^{12}$ Universit{\'e} Libre de Bruxelles, Science Faculty CP230, B-1050 Brussels, Belgium \\
% ...
$^{17}$ Department of Physics, Loyola University Chicago, Chicago, IL 60660, USA \\
$^{18}$ Dept. of Physics and Astronomy, University of Canterbury, Private Bag 4800, Christchurch, New Zealand \\
% ...
$^{22}$ Niels Bohr Institute, University of Copenhagen, DK-2100 Copenhagen, Denmark \\
$^{23}$ Dept. of Physics, TU Dortmund University, D-44221 Dortmund, Germany \\
% ...
$^{26}$ Erlangen Centre for Astroparticle Physics, Friedrich-Alexander-Universit{\"a}t Erlangen-N{\"u}rnberg, D-91058 Erlangen, Germany \\
% ...
$^{40}$ Dept. of Physics and Wisconsin IceCube Particle Astrophysics Center, University of Wisconsin{\textendash}Madison, Madison, WI 53706, USA \\
% ...
$^{42}$ Department of Physics, Marquette University, Milwaukee, WI, 53201, USA \\
% ...
$^{44}$ Bartol Research Institute and Dept. of Physics and Astronomy, University of Delaware, Newark, DE 19716, USA \\
% ...
$^{63}$ Deutsches Elektronen-Synchrotron DESY, Platanenallee 6, 15738 Zeuthen, Germany  \\
$^{64}$ Institute of Physics, Sachivalaya Marg, Sainik School Post, Bhubaneswar 751005, India \\
% ...

\subsection*{Acknowledgements}

\noindent
The authors gratefully acknowledge the support from the following agencies and institutions:
USA {\textendash} U.S. National Science Foundation-Office of Polar Programs,
U.S. National Science Foundation-Physics Division,
U.S. National Science Foundation-EPSCoR,
Wisconsin Alumni Research Foundation,
Center for High Throughput Computing (CHTC) at the University of Wisconsin{\textendash}Madison,
Open Science Grid (OSG),
Advanced Cyberinfrastructure Coordination Ecosystem: Services {\&} Support (ACCESS),
Frontera computing project at the Texas Advanced Computing Center,
U.S. Department of Energy-National Energy Research Scientific Computing Center,
Particle astrophysics research computing center at the University of Maryland,
Institute for Cyber-Enabled Research at Michigan State University,
and Astroparticle physics computational facility at Marquette University;
Belgium {\textendash} Funds for Scientific Research (FRS-FNRS and FWO),
FWO Odysseus and Big Science programmes,
and Belgian Federal Science Policy Office (Belspo);
Germany {\textendash} Bundesministerium f{\"u}r Bildung und Forschung (BMBF),
Deutsche Forschungsgemeinschaft (DFG),
Helmholtz Alliance for Astroparticle Physics (HAP),
Initiative and Networking Fund of the Helmholtz Association,
Deutsches Elektronen Synchrotron (DESY),
and High Performance Computing cluster of the RWTH Aachen;
Sweden {\textendash} Swedish Research Council,
Swedish Polar Research Secretariat,
Swedish National Infrastructure for Computing (SNIC),
and Knut and Alice Wallenberg Foundation;
European Union {\textendash} EGI Advanced Computing for research;
Australia {\textendash} Australian Research Council;
Canada {\textendash} Natural Sciences and Engineering Research Council of Canada,
Calcul Qu{\'e}bec, Compute Ontario, Canada Foundation for Innovation, WestGrid, and Compute Canada;
Denmark {\textendash} Villum Fonden, Carlsberg Foundation, and European Commission;
New Zealand {\textendash} Marsden Fund;
Japan {\textendash} Japan Society for Promotion of Science (JSPS)
and Institute for Global Prominent Research (IGPR) of Chiba University;
Korea {\textendash} National Research Foundation of Korea (NRF);
Switzerland {\textendash} Swiss National Science Foundation (SNSF);
United Kingdom {\textendash} Department of Physics, University of Oxford.

\end{document}